\documentclass{appolb}
\usepackage{graphicx}
\usepackage{float}
\usepackage[numbers,sort&compress]{natbib}

\usepackage{calc}
\usepackage{color}
\usepackage{xcolor}
\usepackage{verbatim}
\usepackage{float}

\usepackage{enumerate}
\usepackage{fancybox}
\usepackage{amsmath}

\usepackage{lipsum}
\usepackage{afterpage}
\usepackage{xcolor}

\usepackage{subfig}

\begin{document}
\title{Recent results in Euclidean dynamical triangulations%
\thanks{Presented at the 3rd conference of the Polish society on relativity}%
}
\author{J. Laiho, S. Bassler, D. Du, J. T. Neelakanta
\address{Department of Physics, Syracuse University, Syracuse, NY, USA}
\vspace{0.5cm} 
\\{D. Coumbe \thanks{Speaker}}
\address{The Niels Bohr Institute, Copenhagen University, Blegdamsvej 17, DK-2100 Copenhagen, Denmark}
}
\maketitle
\begin{abstract}
We study a formulation of lattice gravity defined via Euclidean dynamical triangulations (EDT). After fine-tuning a non-trivial local measure term we find evidence that four-dimensional, semi-classical geometries are recovered at long distance scales in the continuum limit. Furthermore, we find that the spectral dimension at short distance scales is consistent with 3/2, a value that is also observed in the causal dynamical triangulation (CDT) approach to quantum gravity.
\end{abstract}
  
\section{Introduction}

The perturbative nonrenormalizability of gravity motivates the search for nonperturbative formulations, one example of which is the asymptotic safety scenario~\cite{Weinberg79}. This scenario requires a non-trivial fixed point that the renormalization group flow of gravitational couplings are attracted to at high energies, making gravity effectively renormalizable when formulated nonperturbatively. Although there is currently no rigorous proof for the existence of such a fixed point, the accumulation of evidence is now strongly suggestive~\cite{Reuter:2001ag,Lauscher:2001ya,Litim:2003vp,Codello:2007bd,Codello:2008vh,Benedetti:2009rx,Ambjorn05,Ambjorn:1998xu,Ambjorn:2008wc}. 


In a lattice formulation of an asymptotically safe field theory, the fixed point would appear as a second-order critical point, the  approach to which would define a continuum limit. The  divergent  correlation length characteristic  of  a  second-order  phase  transition would allow one to take the lattice spacing to zero while keeping observable quantities fixed in physical units.  

Euclidean dynamical triangulations (EDT) is a particularly simple formulation of lattice gravity, and one that has already proved successful in two dimensions~\cite{Ambjorn:2002uk}. However, early EDT simulations in four dimensions encountered a number of problems. In particular, the parameter space was found to consist of only two phases, neither of which resembled semi-classical general relativity in four dimensions~\cite{Ambjorn:1991pq,deBakker:1994zf,Ambjorn:1995dj,Catterall:1994pg,Egawa:1996fu}. Moreover, the phase transition separating the two phases was found to be first order, making the existence of a continuum limit unlikely~\cite{Bialas:1996wu,deBakker:1996zx}. The problems encountered in EDT caused Ambjorn and Loll to introduce a causality constraint, which distinguishes between space-like and time-like links on the lattice and thereby permits an explicit foliation of the lattice into space-like hypersurfaces of fixed topology, a formulation known as causal dynamical triangulations (CDT). It has been shown that CDT possesses a 4-dimensional de Sitter-like phase~\cite{Ambjorn05}. Here, we revisit the 4-dimensional EDT approach including a non-trivial measure term, showing that for a specific fine-tuning of the non-trivial measure we obtain results similar to those found in the CDT approach.   

EDT defines a spacetime of locally flat $d$-dimensional triangles, each with a fixed edge length. The model described in this work uses the partition function

\begin{equation}
Z_{E}={\sum_{T}}\frac{1}{C_{T}}\left[\prod_{j=1}^{N_{2}}\mathcal{O}\left(t_{j}\right)^{\beta}\right]e^{-S_{E}},
\end{equation}

\noindent where the product is over all triangles, and \begin{math}\mathcal{O}\left(t_{j}\right)\end{math} is the order of the triangle \emph{j}, i.e. the number of $4$-simplices to which the triangle belongs. The term in square brackets defines our non-trivial measure term, where $\beta$ is a free parameter. The Einstein-Regge action is given by

\begin{equation}
S_{E}=-\kappa_{2}N_{2}+\kappa_{4}N_{4},
\end{equation}

\noindent where \begin{math}\kappa_{2}\end{math} and \begin{math}\kappa_{4}\end{math} are related to the bare Newton's constant and the cosmological constant, respectively. \begin{math}\kappa_{4}\end{math} is tuned to its critical value such that an infinite volume limit can be taken~\cite{deBakker:1994zf}, leaving a 2-dimensional parameter space spanned by \begin{math}\kappa_{2}\end{math} and \begin{math}\beta\end{math}.

\begin{figure}[H]
  \centering
  \includegraphics[width=0.6\linewidth]{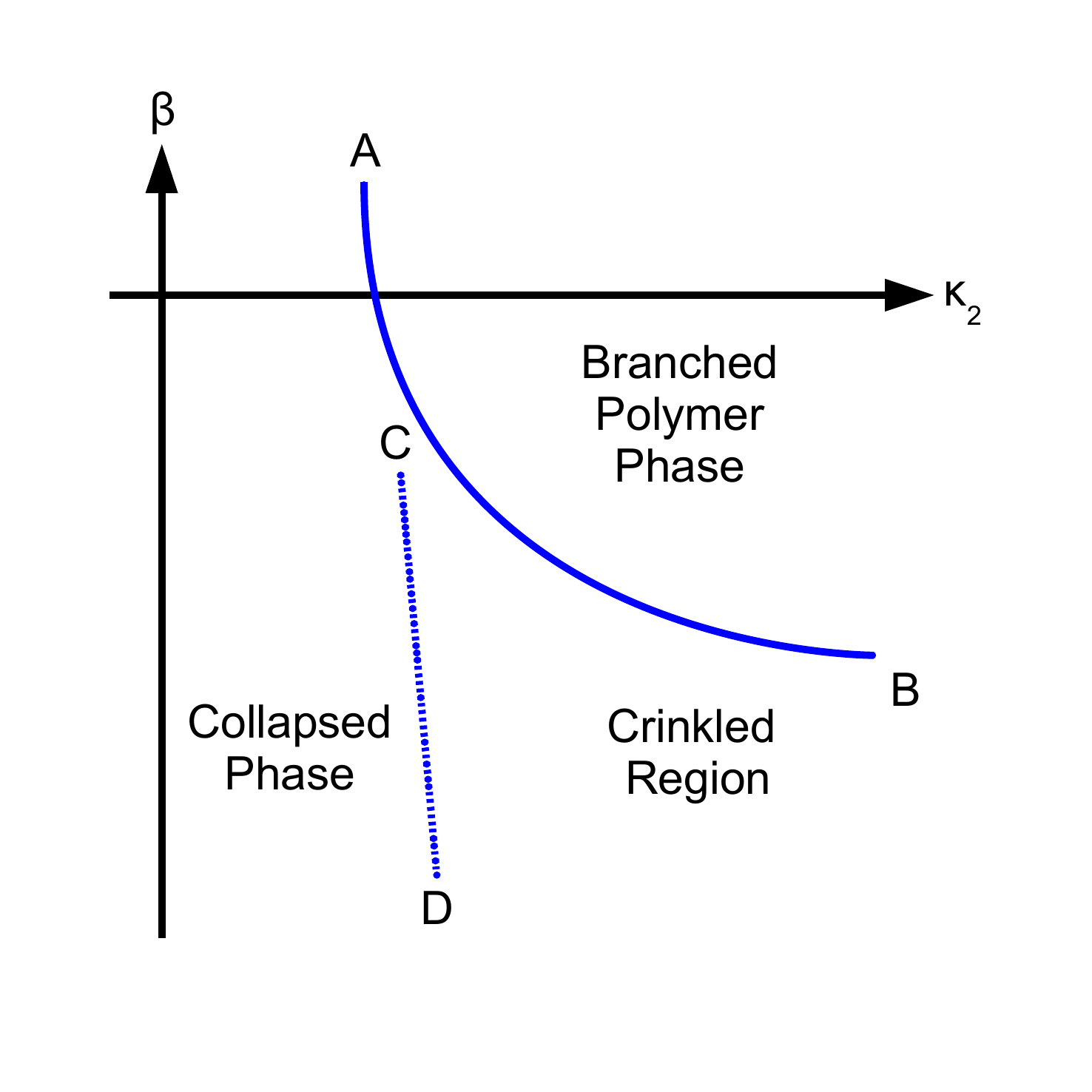}
  \caption{\small A schematic of the EDT phase diagram as a function of $\kappa_{2}$ and $\beta$.}
\label{PD}
\end{figure}

The parameter space of our model is depicted schematically in Fig~\ref{PD}. Numerical simulations have established that the transition line $AB$ is first order and the line $CD$ is a higher-order transition or analytic crossover. In this work we determine the Hausdorff and spectral dimension close to the transition line $AB$.

\section{Global Hausdorff dimension}

The Hausdorff dimension generalises the definition of dimension to non-integer values and can be used to characterise a fractal geometry. We determine the Hausdorff dimension of our ensembles by studying the finite-size scaling of the three-volume correlator $C_{N_{4}}\left(\delta\right)$ introduced in Ref.~\cite{Coumbe:2014nea}, where

\begin{equation}
C_{N_{4}}\left(\delta\right)=\sum_{\tau=1}^t\frac{\left\langle N_{4}^{\rm shell}(\tau)N_{4}^{\rm shell}(\tau+\delta)\right\rangle }{N_{4}^{2}}.
\end{equation}

\noindent $N_{4}^{shell}(\tau)$ is the number of $4$-simplices within a spherical shell one $4$-simplex thick at a geodesic distance $\tau$ from a randomly chosen origin, and $t$ is the maximum number of shells in the triangulation. $N_{4}$ is the total number of $4$-simplices and the normalization of the correlator is chosen such that \begin{math}\sum_{\delta=0}^{t-1}C_{N_{4}}\left(\delta\right)=1\end{math}. We rescale \begin{math}\delta\end{math} via \begin{math}x=\delta/N_{4}^{1/D_{H}}\end{math}, which allows us to determine $D_{H}$ as the value for which \begin{math}c_{N_{4}}\left(x\right)=N_{4}^{1/D_{H}}C_{N_{4}}\left(\delta/N_{4}^{1/D_{H}}\right)\end{math} becomes independent of $N_{4}$.

\begin{figure}[H]
  \centering
  \includegraphics[width=0.6\linewidth]{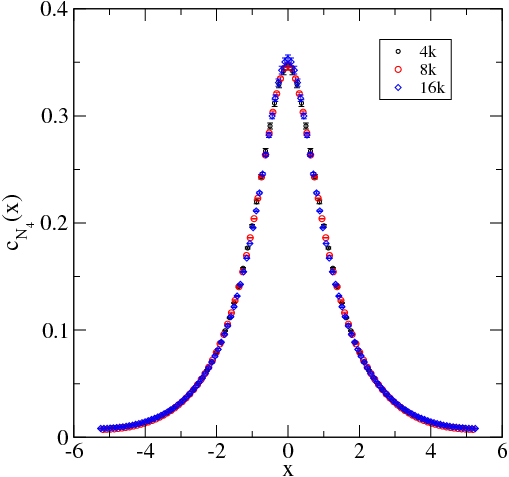}
  \caption{\small Scaling of the volume-volume distribution as a function of the rescaled variable $x=\delta /N_{4}^{1/D_{H}}$ using lattice volumes of 4K, 8K and 16K four-simplices.}\label{Haus}
\end{figure}

  Figure~\ref{Haus} shows the correlator $c_{N_{4}}\left(x\right)$ for lattice volumes of 4K, 8K and 16K four-simplices on the transition line $AB$ for $\beta=0$. We find that the overlap between the curves is maximised for $D_{H}=4.1 \pm 0.3$, providing strong evidence that the Hausdorff dimension close to the transition line $AB$ is consistent with four.  

\section{Spectral dimension}


The spectral dimension $D_{S}$ defines the effective dimension of a fractal geometry via a diffusion process. $D_{S}$ is related to the probability of return $P_{r}$ for a random walk over an ensemble of triangulations after $\sigma$ diffusion steps, and is defined via

\begin{equation}
D_{S}\left(\sigma\right)=-2\frac{d\rm{log}\langle P_{r}\left(\sigma\right)\rangle}{d\rm{log}\sigma}.
\end{equation}



\begin{figure}[H]
\begin{minipage}{.5\linewidth}
\centering
\subfloat[]{\label{main:a}\includegraphics[scale=.35]{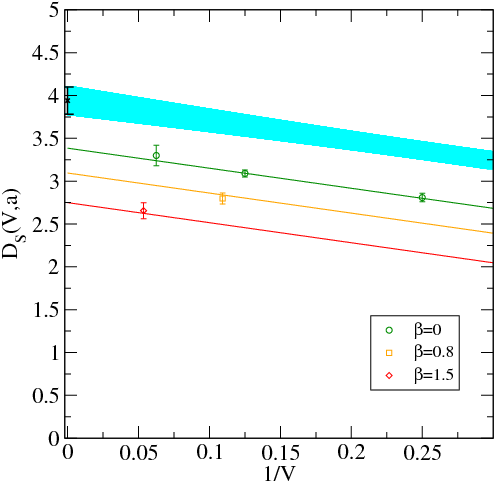}}
\end{minipage}%
\begin{minipage}{.5\linewidth}
\centering
\subfloat[]{\label{main:b}\includegraphics[scale=.35]{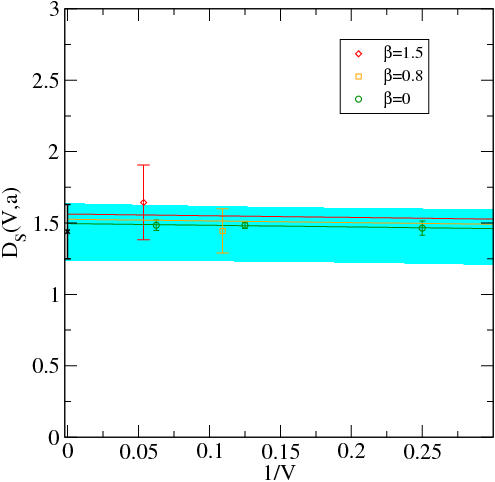}}
\end{minipage}
\caption{The large (a) and small (b) distance scale spectral dimension $D_{S}$ as a function of inverse lattice volume for 3 different $\beta$ values, including an extrapolation to the infinite volume and continuum limit.}
\label{fig:main}
\end{figure}

Assuming the fit function $D_{S}\left(\sigma\right)=a-\frac{b}{c+\sigma}$ we obtain a large distance spectral dimension in the range $D_{S}=2.7-3.3$~\cite{Laiho:2016nlp}, which is inconsistent with 4-dimensional semi-classical general relativity. However, this discrepancy may be due to finite volume or discretisation effects associated with the lattice simulations. In order to investigate whether this is the case we consider an additional extrapolation of $D_{S}(\infty)$ of the form

\begin{equation}
D_{S}(\infty)=c_{0}+c_{1}\frac{1}{V}+c_{2}a^{2},
\end{equation}

\noindent where $c_{i}$ is a fit parameter, $V$ is the volume and $a$ the lattice spacing. This particular ansatz is motivated by the fact that the data points are linear in $1/V$ and $a^{2}$. Extrapolation to the continuum and infinite volume limit gives $D_{S}(\infty)=3.94 \pm 0.16$ and $D_{S}(0)=1.44 \pm 0.19$, as shown in Fig.~\ref{main:a} and Fig.~\ref{main:b}, respectively. A value of $D_{S}(0)$ consistent with 3/2 may have important implications for the asymptotic safety scenario~\cite{Laihobb}.  

\section{Discussion and conclusions}


In this work we determine the Hausdorff and spectral dimension for a specific fine-tuning of the bare coupling constants in Euclidean dynamical triangulations (EDT). Using a finite-size scaling analysis we determine the Hausdorff dimension to be $D_{H}=4.1\pm 0.3$ on the transition line $AB$ for $\beta=0$, which is consistent with 4-dimensional general relativity and CDT results~\cite{Ambjorn05}. Furthermore, by applying an additional extrapolation to the continuum and infinite volume limits we find a large scale spectral dimension of $D_{S}(\sigma)=3.94 \pm 0.16$ and a small distance value of $D_{S}(\sigma)=1.44 \pm 0.19$, results that are also similar to those reported in CDT~\cite{Ambjorn:2005db,Coumbe:2014noa}.







\bibliographystyle{unsrt}
\bibliography{Master}



\end{document}